\documentclass{cup-hpl}

\usepackage{graphicx}
\usepackage{caption}
\usepackage{subcaption}

\usepackage{titlesec}
\titleformat*{\paragraph}{\itshape\mdseries}

\usepackage{easyReview}
\setreviewson
\usepackage{gensymb}

\usepackage{lipsum}
\usepackage{units}
\usepackage{todonotes}
\setuptodonotes{inline}
\usepackage[hidelinks]{hyperref} 
\usepackage[capitalize]{cleveref}

\begin{document}

\shorttitle{Two-chamber gas target for laser-plasma accelerator electron source}                                   

\shortauthor{Drobniak et al.}

\title{Two-chamber gas target for laser-plasma accelerator electron source}


\author[1]{P. Drobniak}
\author[1]{E. Baynard}
\author[1]{K. Cassou}
\author[1]{D. Douillet}
\author[1]{J. Demailly}
\author[1]{A. Gonnin}
\author[1]{G. Iaquaniello}
\author[1]{G. Kane}
\author[1]{S. Kazamias}
\author[1]{N. Lericheux}
\author[1]{B. Lucas}
\author[1]{B. Mercier}
\author[1]{Y. Peinaud}
\author[1]{M. Pittman}
\address[1]{Laboratoire de Physique des 2 Infinis Irène Joliot-Curie - IJCLab - UMR9012 - Bât. 100 - 15 rue Georges Clémenceau 91405 Orsay cedex - France}
 
\corresp{drobniak@ijclab.in2p3.fr, cassou@ijclab.in2p3.fr}

\date{\today}

\begin{abstract}
Exploring new target schemes for laser wakefield accelerators is essential to meet the challenge of increasing repetition rates while ensuring stability and quality of the produced electron beams. The prototyping of a two-chamber gas cell integrated into the beam line and operating in continuous gas flow is introduced and discussed in the frame of ionisation injection. We report the numerical fluid modeling used to assist the density profile shaping. We describe the test bench used for cell prototype assessment, in particular the plasma electron density and longitudinal distribution of species relevant for ionisation injection. The lifetime of the target key part is measured for different materials. Perspectives to high power operation are outlined. 

\end{abstract}

\maketitle

\section{\label{sec:introduction} Introduction}

Laser wakefield acceleration (LWFA) is a promising high-gradient accelerator technology, and the interest of the accelerator community is growing due to its compactness \cite{assmann2020eupraxia,kit,antipov2021}. Significant progress has been made in the optimisation of laser-plasma electron source (so-called 'target') achieving GeV-level\cite{gonsalves2020}, but also controlled high charge beams, and optimisation of spectral brightness \cite{jalas2021bayesian,kirchen2021optimal}. Long operation runs at various repetition rates is also a key issue \cite{maier2020,rovige2021}. All these improvements are possible only with advanced control of both laser and plasma target. In the under-dense plasmas used in plasma wakefield accelerators, the gas typically takes the form of supersonic jets, gas cells, capillary discharge waveguides \cite{leemans2006} or plasma ovens \cite{awake2018}. 
Depending on repetition rate and integration constraints, targets are operated in pulsed or continuous gas flow mode. A deep understanding and control of the target density profile, species distribution and gas flow is essential to ensure high-quality and reproducible electron beam production. 

A high compactness approach using a two-chamber target directly integrated into the beamline is developed. Section~\ref{sec:targets} presents a review of existing laser-driven accelerator targets. Section~\ref{sec:cell} introduces the prototype mechanical design with fluid  simulations, together with predicted density profiles. Section~\ref{sec:test} describes the test bench used for target prototype experimental characterisation. Eventually section~\ref{sec:experimental} concludes with the qualification of the fluid simulation model, and prototype lifetime consideration.


\section{\label{sec:targets}Targets for laser-driven plasma accelerator}

As reviewed by I. Prencipe \textit{et al.}\cite{prencipe2017} and J. Garland \textit{et al.}\cite{garland2020plasma}, several plasma target designs have been investigated in the last two decades: mainly gas jets, gas cells and capillary discharges. In all designs tried, the challenge is to tune the plasma composition and longitudinal density profile. For the particular case of laser-driven electron injectors, the target is composed by a first stage where injection occurs, a second stage for acceleration and a third with controlled density ramp to limit emittance growth\cite{lehe2014}. The various approaches are summarised in Tab.\ref{tab:targetTypeCharacterisics}.

Gas jets are the most commonly used and often based on a single jet technique using either the principle of self-injection \cite{faure2004laser}, optical injection (with colliding pulses \cite{faure2006controlled}), ionisation injection \cite{pak2010injection,mcguffey2010ionization}, or down-ramp injection \cite{chien2005spatially, brijesh2012tuning, buck2013shock, burza2013laser}  triggered by a shock using a blade \cite{chien2005spatially, buck2013shock} or a wire \cite{burza2013laser} or by shaping the plasma with a transverse beam \cite{brijesh2012tuning}. Other schemes have been proposed using two jets, the first jet being the injector, the second one the accelerating stage. For the injection, the techniques tried were down-ramp \cite{hansson2015down} or ionisation \cite{golovin2015tunable} injection. The main advantage of gas jets is the easy alignment with the laser and the wide solid angle for diagnostics. Pulsed operation is advised to avoid too much gas leaks, leading to pumping system overload and pollution.
At high operation rate (typically kHz), gas jet high density tends to induce high thermal and mechanical loads, resulting in wearing of mechanical parts, vibrations and shot-to-shot instability \cite{garland2020plasma, gustas2018high, guenot2017relativistic}.
Typical electron densities offered by gas jets lie in the range of $10^{18}-10^{20}$~cm$^{-3}$.

\begin{table*}[tp!]
    \centering
    \begin{tabular}{|l|c|c|c|c|}
         \hline
         target type & density range [cm$^{-3}$] & distribution species & lifetime [shots] & repetition rate [Hz]   \\
         \hline 
         jet &  $10^{18}-10^{20}$ & mixed & $>10^5$ & $0.1-1000$  \\
         cells &  $10^{17} - 10^{19}$ & mixed, localised & $>10^{5}$ & $0.1-10$  \\
         channels & $10^{17} - 5\times10^{18}$ & mixed,  localised & $>10^{4}$ & $0.1-10$  \\
         capillary discharge & $10^{17} - 5\times10^{18}$ & mixed, localised & $>10^{4}$ & $0.1-10$\\
         \hline
    \end{tabular}
    \caption{Overview of state-of-the-art target properties for laser-plasma accelerator electron source}
    \label{tab:targetTypeCharacterisics}
\end{table*}

Gas cells are divided into two categories. The first one is a tank \cite{audet2018gas, dickson2022mechanisms} or several tanks \cite{pollock2011demonstration, kononenko20162d}, filled with gas in steady state flow or pulsed mode. Apertures allow the laser to pass through, and keeping them as small as possible is critical to prevent leaks. The second category is gas channels \cite{desforges2014dynamics, jalas2021bayesian, kirchen2021optimal, kim2021development}, where gas is injected by various transverse inlets into a main longitudinal channel with reduced cross section. In most cases, the first transverse inlet is for electron injection, the other ones for electron acceleration. Gas exhaust occurs at the main channel entrance/exit and may additionally go through a specific transverse aspiration outlet. 

Whether using a tank or channel geometry, gas cells are particularly interesting for an ionisation injection regime, where a fraction of high-Z gas (called dopant) is added into a background gas. 
Various techniques have been developed to avoid continuous ionisation injection using a downward focusing in gas jet or gas cell or 
a sharp confinement of the dopant\cite{pollock2011demonstration}  allowing to reduce the accelerated beam energy spread and control beam-loading \cite{kirchen2021optimal}.

Whereas gas jets require quite high backing pressures (in the range of several bars), gas cells are less demanding in terms of gas consumption and pressure gradient in the gas circuit. Depending on the vacuum integration (differential pumping), they can be operated in pulsed or continuous gas-injection mode, which yields a better shot-to-shot stability \cite{garland2020plasma}. On balance, the main drawback of gas cells are: (1) lifetime, since laser may enlarge cell apertures, (2) reduced solid angle for diagnostics, since their material may diffuse light, or potentially be coated by plasma pollution.



The design investigated here is a gas cell divided in two separate chambers, delimited by transparent optical quality plane surfaces, specifically suited for transverse optical diagnostics. It is inspired by the pioneer work done by Kononenko \cite{kononenko20162d}, in a more compact approach and focusing on the dopant mitigation in the first zone, with pure background gas in the second zone. 


\section{\label{sec:cell}Target multi-cell design} 

The motivation of this work is: (1) a compact integration directly into the accelerator beam line,
(2) a large range online tunability of dopant concentration and gas density profiles, (3) together with their online transverse optical diagnostics,
(4) an easy replacement of critical elements which are strongly irradiated by the laser, especially at high repetition rate.

\subsection{\label{ssec:design}Design and features}

\begin{figure}[htbp]
        \centering
        \includegraphics[width=0.4\textwidth]{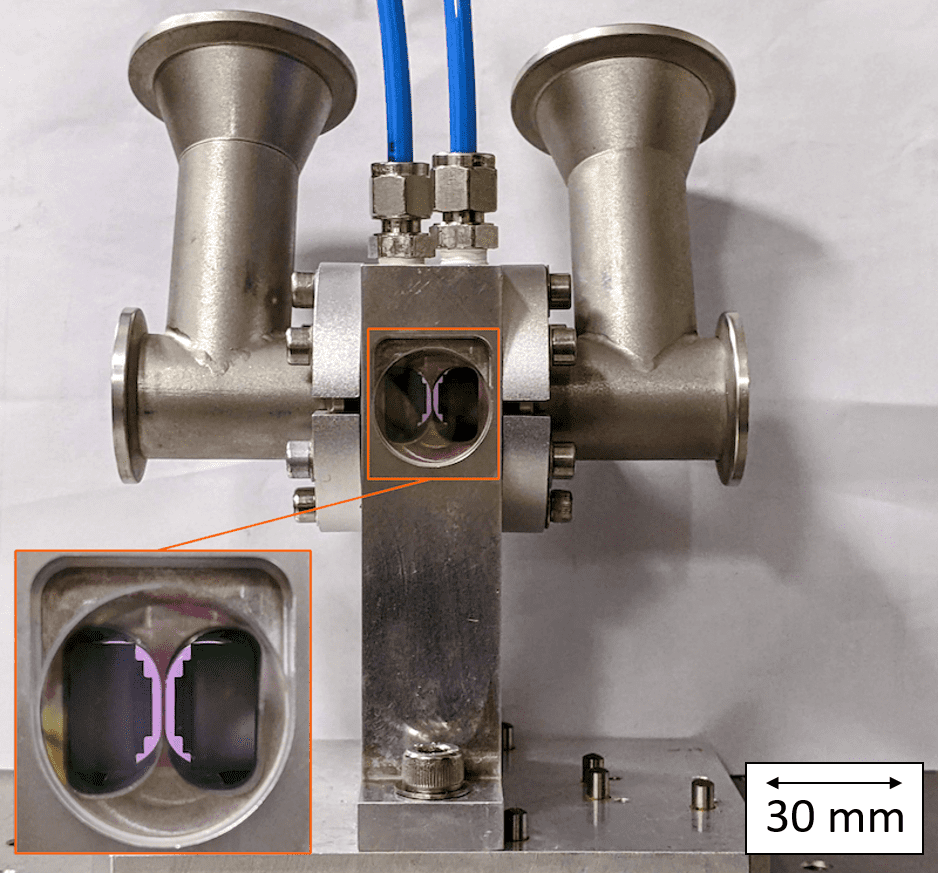}
        \caption{View of the two-chamber target prototype with the input and exit T-pipe for efficient gas evacuation with a zoom on the chambers. The present prototype is equipped with MACOR ceramic nozzle.}
        \label{fig:cellPicture}
    \end{figure}

\begin{figure}[htbp]
        \centering
        \includegraphics[width=0.4\textwidth]{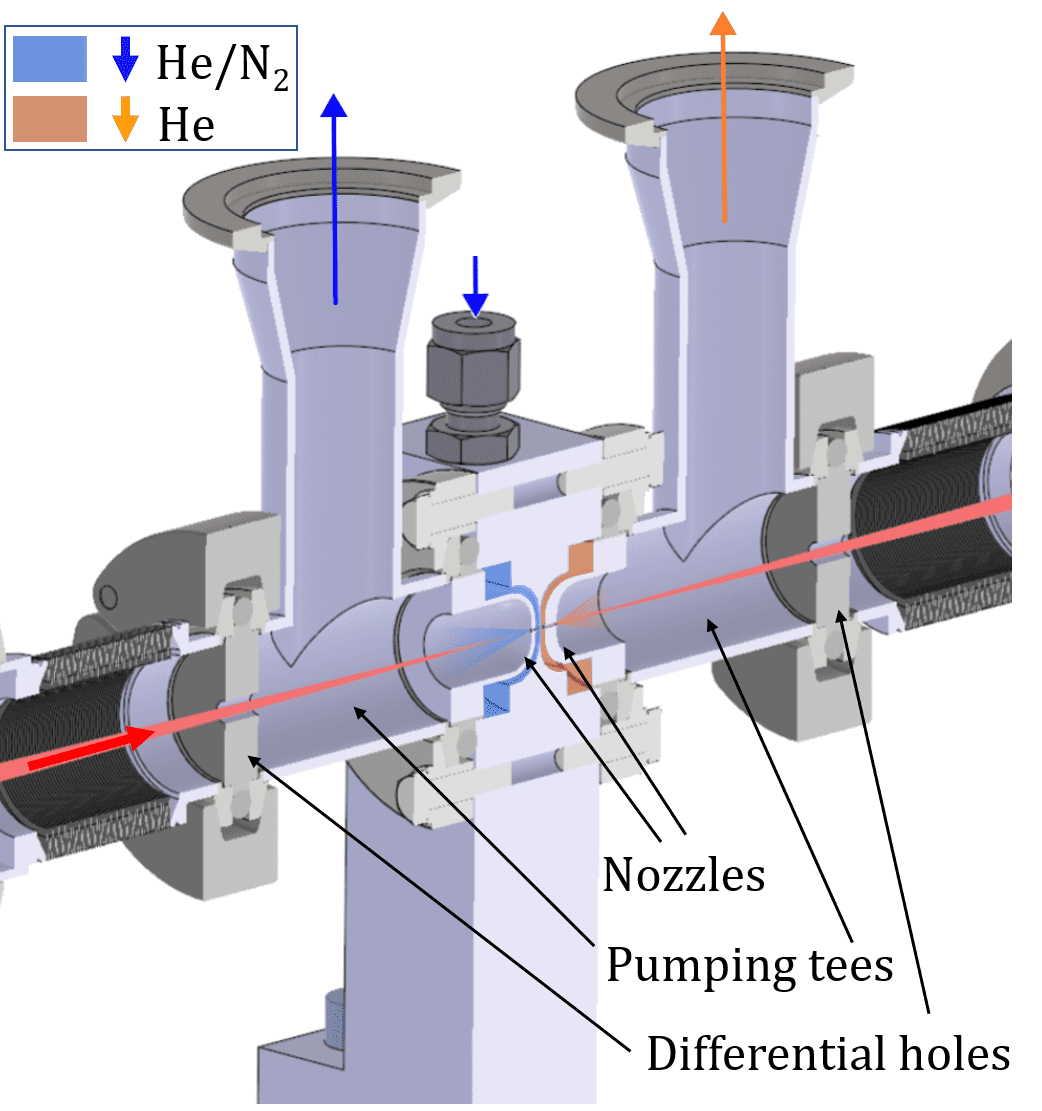}
        \caption{CAD section view of target connected in the beamline with a laser propagating from left to right. Gases are injected with two connections on the top (only one visible in this 'section' view), flow towards the chambers (blue and orange areas), and exit through the nozzles. They expand in the pumping tees and most of the flow is sucked out by efficient primary pumping (upwards in this schematic view). Two $8\,$mm-apertures (differential holes) provide a differential pumping at the entrance and exit of the T-pipes.}
        \label{fig:cellScheme_screenshot}
    \end{figure}

The prototype design is presented in Fig.~\ref{fig:cellPicture} and Fig.~\ref{fig:cellScheme_screenshot}. It consists in a main body and two nozzles defining two separate chambers (called chamber 1 and 2 along laser propagation direction), each supplied with gas through an injection hose: helium doped with nitrogen ($He$/$N_2$) for the chamber 1 and pure helium ($He$) chamber 2. They are separated by a wall, with a small central aperture ranging from $0.25\,$ to $1\,$mm in diameter. The laser enters chamber 1 through the inlet nozzle, passes through the central aperture and exits chamber 2 by the outlet nozzle. 

Nozzles are necessary to reduce the leak in the vacuum chamber. Their inner dimensions (radius, thickness) allow an additional tuning of the gas density in- and out-ramp shape. The central separation serves as frontier between the target chambers. Some gas flow between chambers may appear and is governed by the pressure difference and conductance of the central aperture. 

\begin{figure}[htbp]
        \centering
        \includegraphics[width=0.4\textwidth]{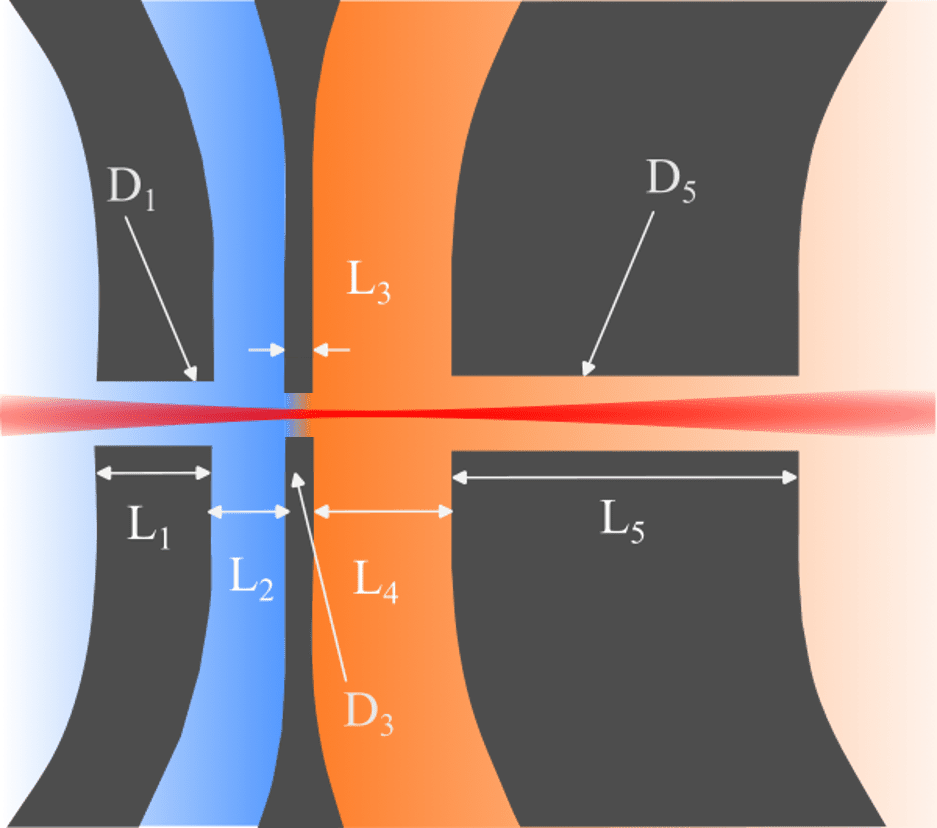}
        \caption{Cell dimensions nomenclature with associated variables. $D_1$, $D_3$ and $D_5$ respectively are the diameters of: inlet nozzle, central aperture and outlet nozzle. $L_1$, $L_2$, $L_3$, $L_4$ and $L_5$ respectively correspond to the lengths of: inlet nozzle, chamber 1, central aperture, chamber 2 and outlet nozzle. Gas paths are indicated for $He/N_2$ (blue filled region) and $He$ (orange filled region). Laser path is schemed in red and propagates from left to right.}
        \label{fig:cellDimensions}
    \end{figure}

Dimensions of the cell close to the axis are described in Fig.~\ref{fig:cellDimensions} and given in Tab.~\ref{tab:dimensions}. Many combinations have been considered, and a typical configuration is given in Tab.~\ref{tab:dimensions}. Varying the nozzle total length gives an adjustment of chamber 1 and 2 longitudinal dimensions called $L_2$ and $L_4$ (see Fig.~\ref{fig:cellDimensions}). 
The tank volume of each chamber is $\sim 5\,$cm$^{3}$.

    \begin{table}[htbp]
    \centering
    \caption{Typical cell dimensions close to the axis, with nomenclature defined in Fig.~\ref{fig:cellDimensions}. All in mm.}
    \begin{tabular}{c|c|c|c|c|c|c|c}
        $L_1$ & $L_2$ & $L_3$ & $L_4$ & $L_5$ & $D_1$ & $D_3$ & $D_5$ \\
        \hline
        $1$ & $0.6$ & $0.25$ & $1.2$ & $3$ & $0.6$ & $0.25$ & $1$ \\
    \end{tabular}
    \label{tab:dimensions}
    \end{table}

Primary vacuum (sub-mbar) is ensured close to the nozzle exit by a pumping system connected with T-pipes (Fig.~\ref{fig:cellScheme_screenshot}). Secondary vacuum is obtained further from the cell after a differential hole, both downstream and upstream, which produces a two-decade pressure drop.

The main body has been manufactured using wire electro-discharge machining in aluminium block, while nozzles are either made of aluminium or MACOR ceramics. 
The centering mechanical tolerance  ($\pm 50\,\mu$m center to center) was achieved for the nozzles using numerical milling machining.  

In addition to its mechanical features, the design allows to perform transverse optical diagnostics across chamber 1 and 2. The diagnostics can be placed in air thanks to optical windows, that are the direct frontier between chambers and experimental room. Such a feature is particularly interesting for convenient experimental measurements of gas and plasma characteristics. Compared to channel type gas cells the optical transverse diagnostics are eased even if the central separation wall introduces shadowing in the imaging of the two chambers for 2D spectroscopic light collection. The transverse distance from the center (interaction region) and the optical windows is $\approx 3\,$cm avoiding a rapid darkening due to  pollution by the laser.

\subsection{\label{ssec:fluid} Fluid simulation set-up }


The gas density distribution is modelled using the open-source fluid simulation code OpenFOAM \cite{openfoam}. Typical simulation cases from this article are online and open to the scientific community \cite{git}. Depending on the desired problematic, the solver used is either:
\begin{itemize}
    \item \textit{rhoPimpleFoam} \cite{rhopimplefoam}: for transient compressible single species simulation,
    \item \textit{interMixingFoam} \cite{intermixingfoam}: for transient incompressible miscible fluids,
\end{itemize}
No solver modeling miscibility for compressible flows were found in the OpenFOAM library, therefore two simulation steps were necessary.

\begin{figure}[htbp]
        \centering
        \includegraphics[width=0.5\textwidth]{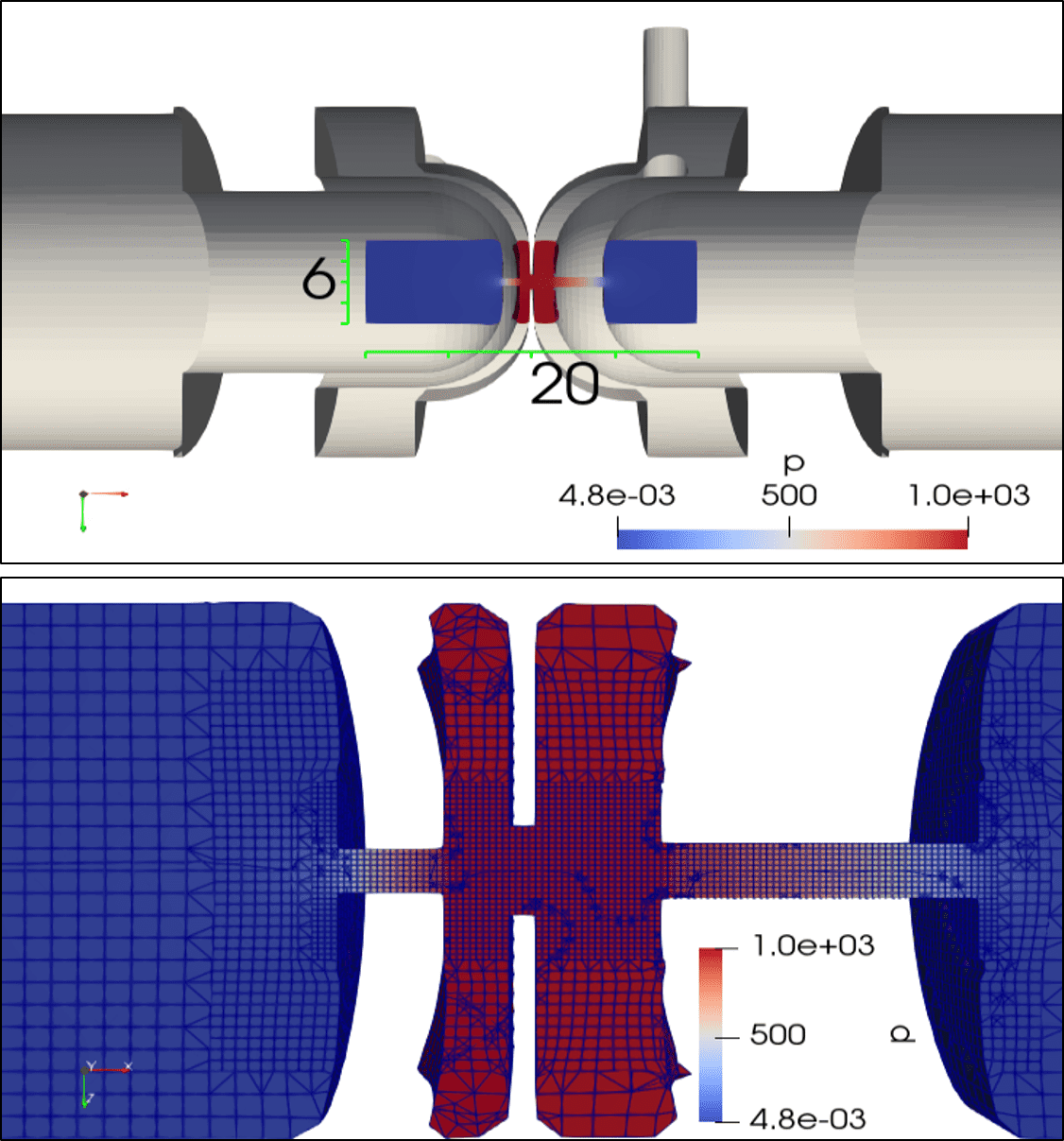}
        \caption{Longitudinal clip of the reduced mesh used in OpenFOAM simulations for a target configuration with geometry $(L_{1},L_{2},L_{3},L_{4},L_{5},D_{1},D_{3},D_{5})=(1,0.6,0.25,1.2,3,0.5,0.95,0.6)$. The \textit{snappyHexMesh}-generated mesh is presented in the original CAD design .stl file (top image, the .stl is in grey) and zoomed-in with visible mesh refinement areas (bottom image). $p$ is the pressure in Pa. Laser travels from left to right.}
        \label{fig:OF_mesh_reduced}
    \end{figure}

The geometry is designed with a CAD software and automatic meshing is performed using the routine \textit{snappyHexMesh} \cite{snappyhexmesh}.
A 3D geometry applied to a reduced volume is used for simulations, as presented in Fig.~\ref{fig:OF_mesh_reduced}, in order to limit the total number of cells to an average of $10^{5}$ and thus limit the computation time.

\noindent Boundary conditions are: fixed pressure at the inlets, constant volumetric flow at the outlets (estimated from the pump characteristics).



\noindent Simulations are run on a computer single-cpu and the average simulation time is roughly $1\,$h for the reduced volume case.

\subsection{\label{ssec:predicted} Simulation of the density profiles for single species}
\begin{figure}[htbp]
        \centering
        \includegraphics[width=0.5\textwidth]{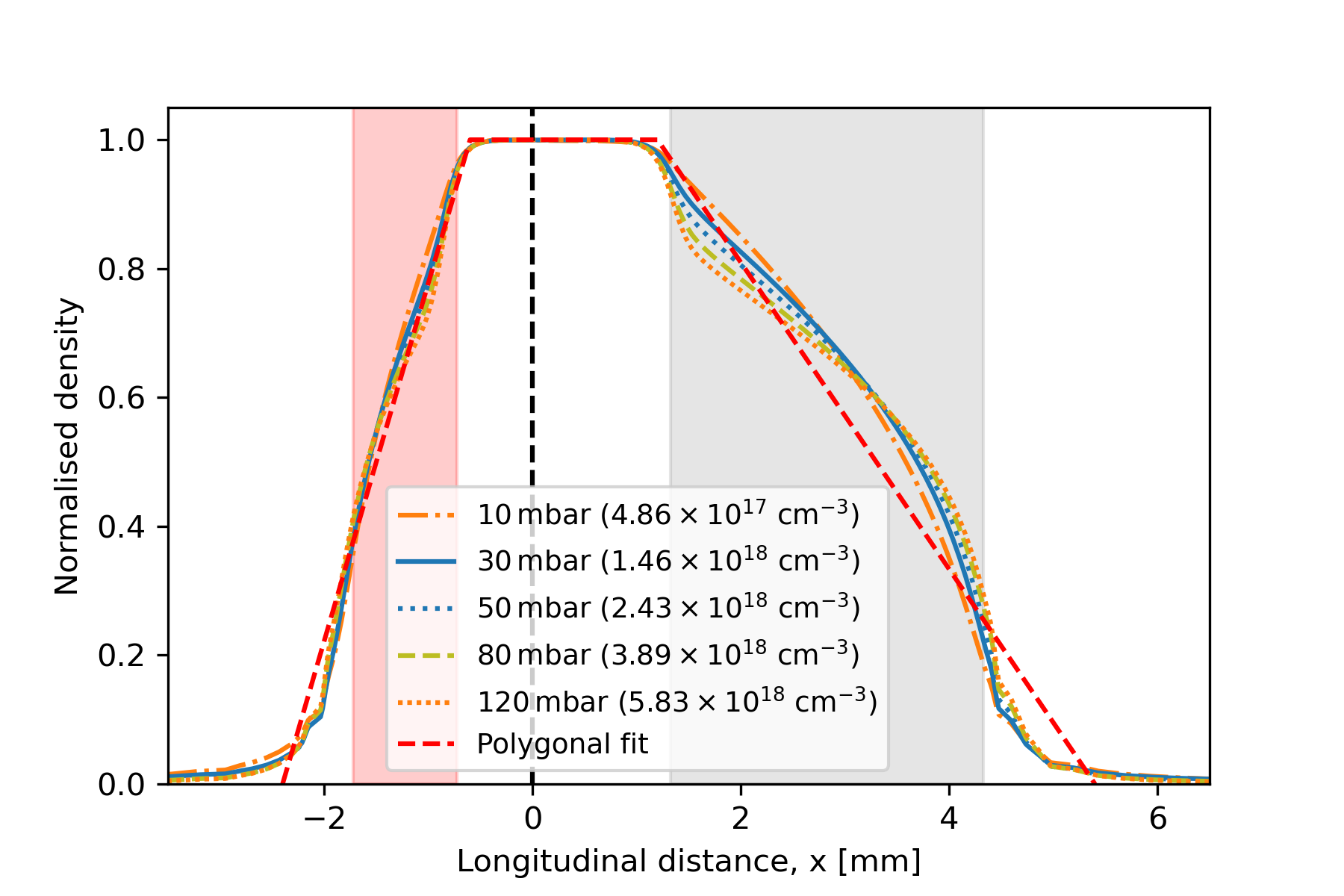}
        \caption{Evolution of longitudinal on-axis density with pressure according to \textit{rhoPimpleFoam} simulations with pure $He$. Simulations are run with $p_{Left} = p_{Right}$ on geometry $(1,0.6,0.25,1.2,3,0.5,0.95,0.6)$, with names referring to the maximum pressure in the plateau. The associated electron density is in parenthesis, assuming $He$ full ionisation. A polygonal fit is added in red. Inlet and outlet nozzle extensions are respectively depicted in red and grey areas. The cell center (central aperture) is depicted with a vertical black dashed line. Laser goes from left to right.}
        \label{fig:OF_plot_He_evolutionWithPressure}
    \end{figure}

First simulations are run for pure $He$ with \textit{rhoPimpleFoam}. The resulting steady-state is obtained from an initial empty cell a few hundreds of $\mu$s after valve opening. The longitudinal density profiles obtained for cell dimensions $(L_{1},L_{2},L_{3},L_{4},L_{5},D_{1},D_{3},D_{5})$ = $(1,0.6,0.25,1.2,3,0.5, \\ 0.95,0.6)$ at several operating pressures are presented in Fig.~\ref{fig:OF_plot_He_evolutionWithPressure}.     
Injection pressures for chamber 1 ($p_{Left}$) and chamber 2 ($p_{Right}$) satisfy $p_{Left} = p_{Right}$, in order to have a flat plateau between both chambers. This feature prevents convection and is further discussed below in the gas mixture simulation. Note that $D_{3}$ is voluntarily taken as $0.95\,$mm to model a $0.25\,$mm realistic damaged central aperture, but simulation results are similar.

Within the $[10;120]$~mbar pressure range, the longitudinal profile shape is conserved, with slight compression effects, mostly at the outlet nozzle entrance. With cylindrical nozzles, the up- and down-ramp shape is pretty linear. The overall pressure profile can thus easily be approximated with scalable linear functions, that can be tabulated over a wide range of pressure to serve as input for optimisation with laser-plasma PIC simulation (see polygonal fit in Fig.~\ref{fig:OF_plot_He_evolutionWithPressure}). Also note that the pressure upstream the inlet nozzle quickly decreases below the mbar range for all configurations, limiting undesired laser-plasma interaction before the cell. 

\begin{figure}[htbp]
        \centering
        \includegraphics[width=0.5\textwidth]{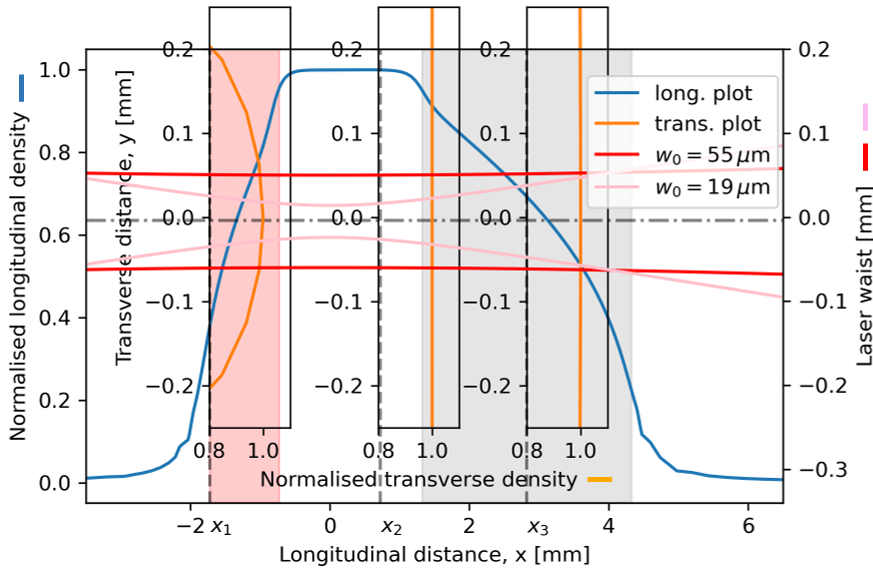}
        \caption{Evolution of transverse density with propagation according to \textit{rhoPimpleFoam} simulations with pure $He$ at injection pressures $p_{Left} = p_{Right} = 30$~mbar, with geometry $(1,0.6,0.25,1.2,3,0.5,0.95,0.6)$. Transverse plots (orange) are extracted at $x_1$, $x_2$ and $x_3$ respectively corresponding to inlet nozzle entrance, chamber 2 center and outlet nozzle center. A longitudinal plot on axis is added (blue). Two typical laser envelopes are added: $w_{0}=55$~$\mu$m in red and $w_{0}=19$~$\mu$m in pink.}
        \label{fig:OF_plot_He_trans}
    \end{figure}

The transverse density profile is depicted in Fig.~\ref{fig:OF_plot_He_trans}, where three planes along the laser propagation axis are selected: inlet nozzle entrance ($x_1$), chamber 2 center ($x_2$), outlet nozzle center ($x_3$). Both chamber pressure is set to $30$~mbar and the geometry is the same as for Fig.~\ref{fig:OF_plot_He_evolutionWithPressure}. 
Fig.~\ref{fig:OF_plot_He_trans} shows a constant transverse density, whether for a $19$~$\mu$m (PALLAS project\cite{pallaswebsite}) or a $55$~$\mu$m (test bench) laser waist at the interface between the two chambers.

\begin{figure}[htbp]
        \centering
        \includegraphics[width=0.5\textwidth]{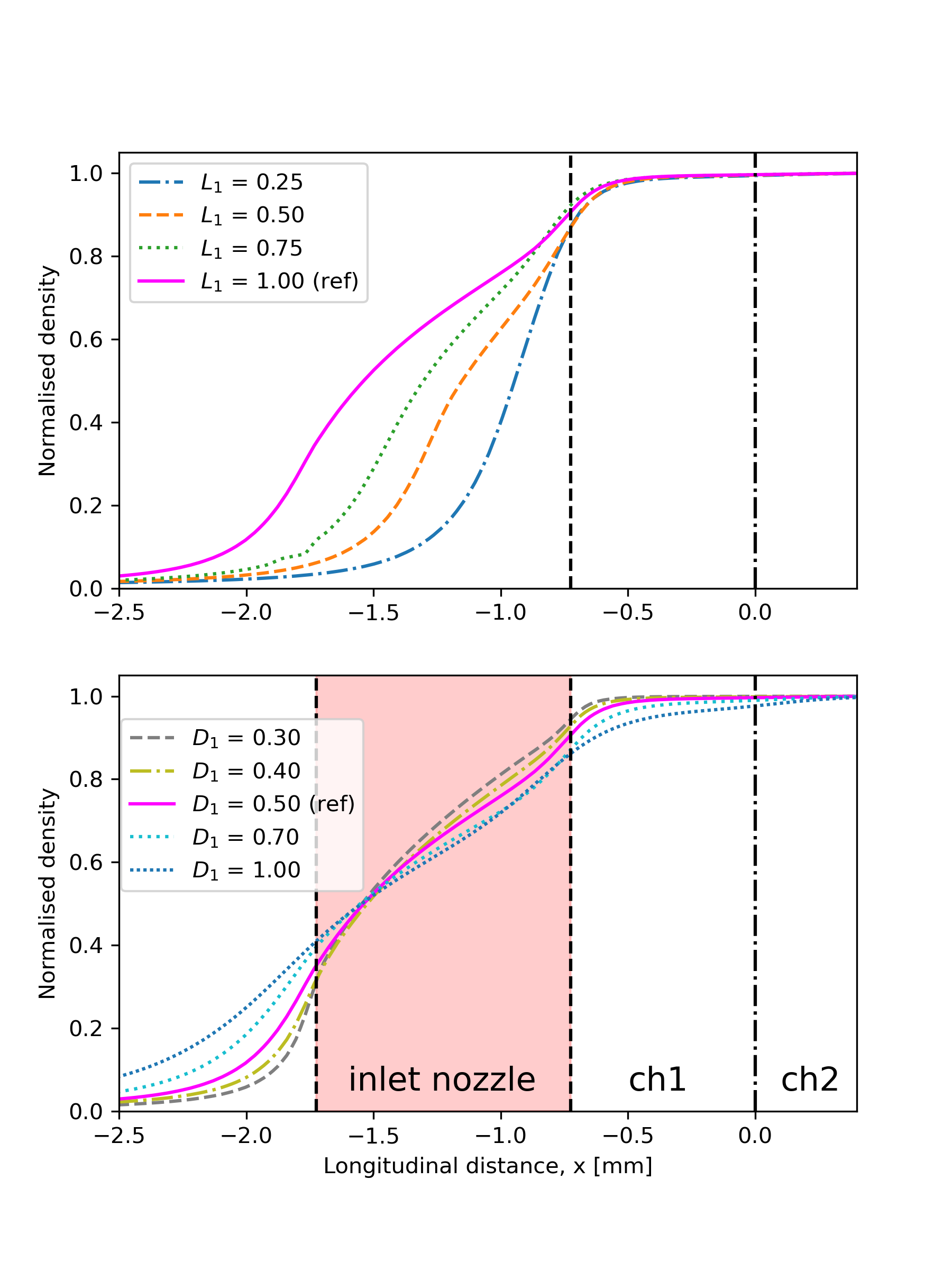}
        \caption{Calculated normalised density for different inlet nozzle geometries from \textit{rhoPimpleFoam} simulations for pure $He$ at $p_{Left}=p_{Right}=30$~mbar. The reference geometry (magenta) is $(L_{1},L_{2},L_{3},L_{4},L_{5},D_{1},D_{3},D_{5})=(1,0.6,0.25,1.2,3,0.5,0.95,0.6)$~mm. The top graph presents results for the inlet nozzle length variation $L_1$: $0.25$, $0.50$, $0.75$ and $1.00$~mm (reference geometry) with constant $D_1=0.50$~mm. The bottom graph shows the influence of inlet nozzle diameter $D_{1}$: $0.30$, $0.40$, $0.50$ (reference geometry), $0.70$ and $1.00$~mm, with constant $L_1=1.00$~mm. Laser travels from left to right.}
        \label{fig:OF_plot_inletNozzle}
    \end{figure}

The influence of inlet nozzle geometry is presented in Fig.~\ref{fig:OF_plot_inletNozzle}, where the reference profile $(1, 0.6, 0.25, 1.2, 3, 0.5, \\0.95, 0.6)$ is kept and compared with other diameter or length.
The longitudinal extent of the up-ramp scales linearly with $L_1$ but does not depend on $D_1$. Increasing the diameter leads to higher upstream tee pressure that reaches the mbar range for $D_1 > 0.5$~mm, but also degrades the flatness of the plateau in chamber 1. Ideally, the shortest and thinnest nozzle as possible is desired. We choose $L_1 = 1$~mm and $D_1 = 0.5$~mm for machining and robustness reasons. $D_1$ must obviously also be larger than a few laser waists.





\begin{figure}[htbp]
        \centering
        \includegraphics[width=0.5\textwidth]{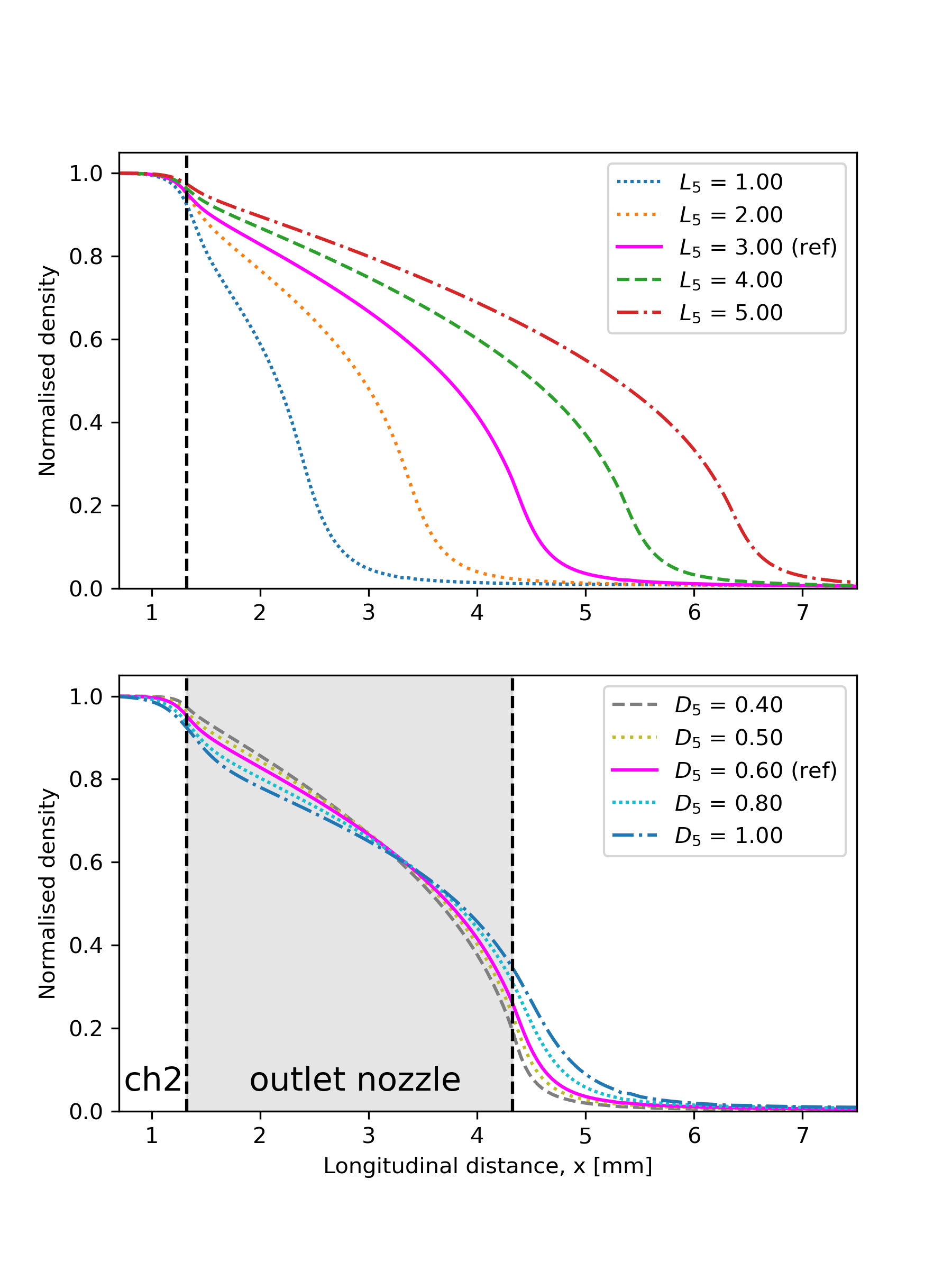}
        \caption{Calculated normalised density on axis for different outlet nozzle geometries from \textit{rhoPimpleFoam} simulations for pure $He$ at $p_{Left}=p_{Right}=30$~mbar. The reference geometry (magenta) is $(L_{1},L_{2},L_{3},L_{4},L_{5},D_{1},D_{3},D_{5})=(1,0.6,0.25,1.2,3,0.5,0.95,0.6)$~mm. The top graph presents results for the outlet nozzle length variation $L_5$: $1.00$, $2.00$, $3.00$ (reference geometry), $4.00$ and $5.00$~mm with constant $D_5=0.60$~mm. The bottom graph shows the influence of outlet nozzle diameter $D_{5}$: $0.40$, $0.50$, $0.60$ (reference geometry), $0.80$ and $1.00$~mm, with constant $L_5=3.00$~mm. Laser travels from left to right.}
        \label{fig:OF_plot_outletNozzle}
    \end{figure}

The influence of outlet nozzle dimensions at $30\,$mbar is presented in Fig.~\ref{fig:OF_plot_outletNozzle}, where the length $L_{5}$ is varied between $[0;5]\,$mm at fixed $D_5 = 0.6$~mm and the diameter $D_{5}$ between $[0.40;1.00]\,$mm at fixed length $L_5 = 3$~mm. The same reference case as for the inlet nozzle study is included for comparison.
Similarly to the inlet nozzle, the ramp length linearly scales with nozzle length $L_5$, with a preserved shape. Indeed, for each length tried (top graph in Fig.~\ref{fig:OF_plot_outletNozzle}), the down-ramp follows the same pattern: a linear decrease along roughly $L_5$ followed by an exponential decrease of $\approx 1$~mm (gas expansion). The outlet diameter $D_5$ has the same influence as previously observed with $D_1$.

Contrary to the inlet, the outlet profile has to be as smooth and long as possible for emittance preservation
\cite{li2019preserving}, which corresponds to large $L_5$.
Together with a small $D_5$, this might be a problem due to laser divergence and possibly ablation.
A compromise is made with  $L_5 = 3\,$mm and $D_5 = 0.60$~mm.

\subsection{\label{ssec:dopant_sim} Simulation of dopant confinement}

As introduced earlier, dopant confinement is a key process to ensure high quality beams with a small energy spread. Specific incompressible two-gas simulations are run with \textit{interMixingFoam} to account for diffusion issues. They are performed in a reduced geometry with boundaries up to each nozzle center, where the flow can still be approximated as incompressible ($Ma<0.3$). The new outlet boundary conditions are simply the pressure values extracted from previous compressible simulations at the new physical boarders.

\begin{figure}[htbp]
        \centering
        \includegraphics[width=0.5\textwidth]{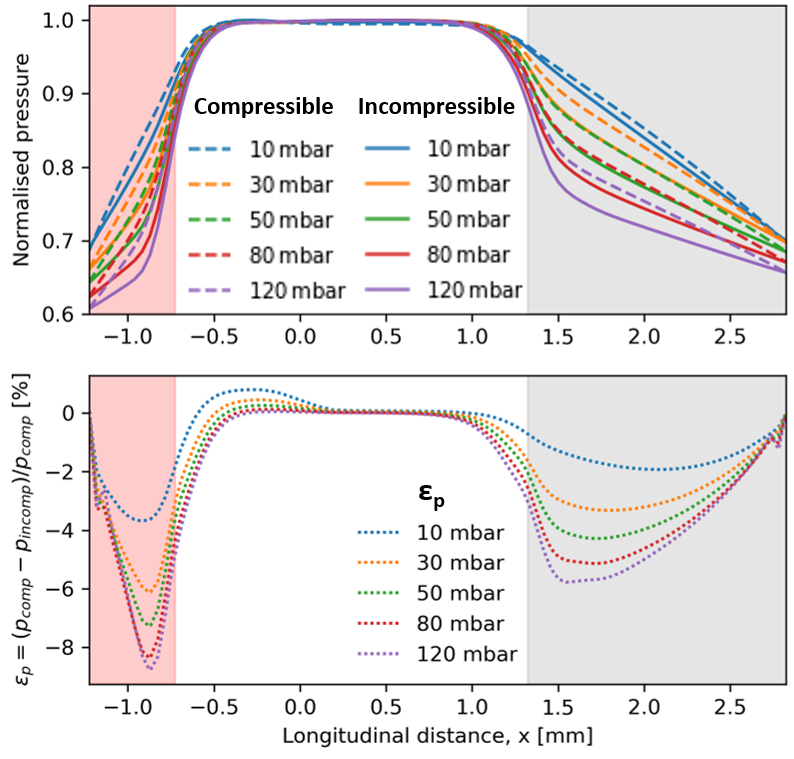}
        \caption{Pressure distribution comparison on axis between compressible (\textit{rhoPimpleFoam}) and incompressible (\textit{interMixingFoam}) simulations, with pure $He$ at plateau pressures: $10$, $30$, $50$, $80$ and $120$~mbar. Upper graph displays compressible plots (dashed line) and incompressible plots (solid line). Lower graph presents the difference between compressible ('comp') and incompressible ('incomp') as: $\epsilon{}_p = (p_{comp}-p_{incomp})/p_{comp}$~[\%]. Target geometry is $(1,0.6,0.25,1.2,3,0.5,0.25,0.6)$ and a reduced mesh is used, limited to the presented x-axis extent. Positions of the inlet and outlet nozzles respectively are indicated by light red and grey areas. Laser goes from left to right.}
        \label{fig:OF_plot_comp_incomp}
    \end{figure}

Such an approximation is verified by simulating comparable cases, both in compressible (\textit{rhoPimpleFoam}) and incompressible (\textit{interMixingFoam}) mode with $He$ only \footnote{Results are still valid when adding a few \% $N_2$ in chamber 1, since the gas characteristics remain comparable, especially in low kinematic areas, such as the chambers interface.}. Results for cell pressures within $[10-120]$~mbar are presented in Fig.~\ref{fig:OF_plot_comp_incomp}, where the difference for the pressure between compressible and incompressible models $\epsilon{}_{p} = (p_{comp}-p_{incomp})/p_{comp}$~[\%] is added. 

From Fig.~\ref{fig:OF_plot_comp_incomp}, a good agreement appears between compressible and incompressible simulations, with a maximum $8\,$\% deviation close to the nozzles, and almost no difference in the chambers, which is the zone where dopant mitigation should occur. The same kind of study was done for temperature and velocity profiles with the same conclusion at the diffusion interface, and a significant divergence close to the nozzles. Incompressible simulations thus correctly match compressible ones \footnote{The reader might note that all plateaus actually have gradients. This particular shape is explained by the difficult and time-consuming search for a perfect match on axis ($p_{Left} = p_{Right}$) using approximate boundary conditions.}.

\begin{figure}[htbp]
        \centering
        \includegraphics[width=0.5\textwidth]{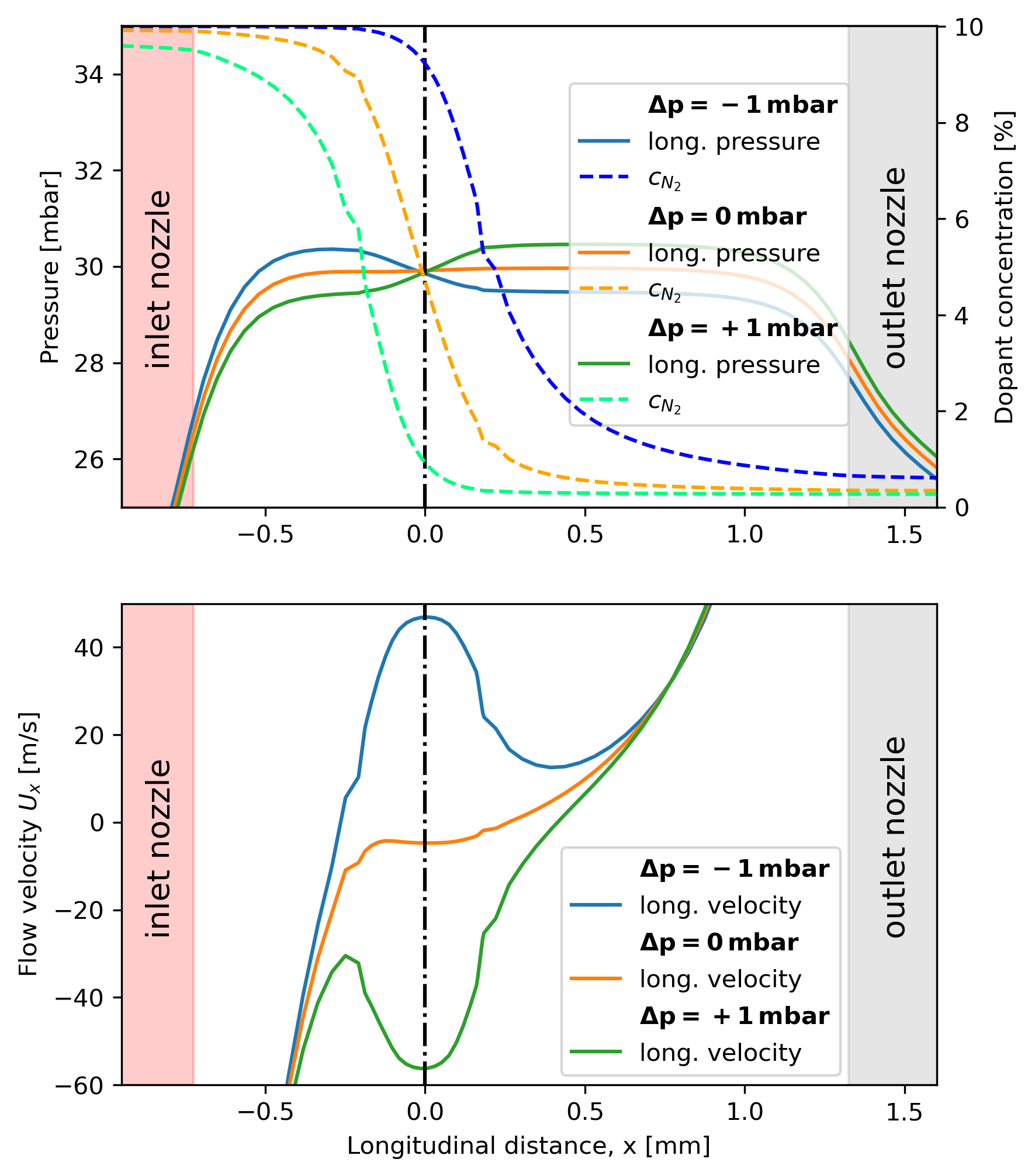}
        \caption{Influence of a pressure gradient between chambers $\Delta{} p = p_{Right}-p_{Left}$ (upper graph, solid lines) on dopant concentration $c_{N_2}$ (upper graph, dashed lines) and longitudinal flow velocity $U_x$ (bottom graph, solid lines). Results are obtained with incompressible miscible simulations (\textit{interMixingFoam}) using $He/N_2$ (at $c_{N_2} = 10\,$\%) and pure $He$ respectively in chamber 1 and 2. Cell geometry used is $(1,0.6,0.25,1.2,3,0.5,0.25,0.6)$ with the same reduced mesh as for Fig.~\ref{fig:OF_plot_comp_incomp}. The cell center is indicated with a vertical dash-dotted line (central aperture) and positions of the inlet/outlet nozzles are added.}
        \label{fig:OF_plot_dopant}
    \end{figure}

The dopant confinement study is then performed with two gases in \textit{interMixingFoam} to evaluate the effect of pressure difference between chamber 1 and 2 $\Delta{} p = p_{Right}-p_{Left}$ (convection) or statistical mixing of the two gases at equal pressure (diffusion). The result is shown in Fig.~\ref{fig:OF_plot_dopant} for a test at $30\,$mbar, with $He/N_2$ injected in chamber 1 at dopant concentration $c_{N_2}=10\%$ and pure $He$ in chamber 2.

For a negative/positive gradient $\Delta{} p = -1/+1$~mbar, the dopant is pushed to the right/left through convection (visible on the flow velocity $U_x$ in Fig.~\ref{fig:OF_plot_dopant}). $\Delta{} p = -1$~mbar causes $N_2$ leaks towards chamber 2 and $c_{N_2}$ never reaches $0$ in chamber 2 (no dopant confinement). $\Delta{} p = +1$~mbar triggers the opposite effect, with $He$ leaking into chamber 1. This case however offers $c_{N_2} = 0$ in chamber 2 (dopant confinement). In both cases, the transition from $c_{N_2} = 10\,$\% to roughly $0$ occurs on $\approx 1.0$~mm.

For equal pressures, the interface is centered on the central aperture ($x=0$) and the $c_{N_2}$ transition is due to pure diffusion (no longitudinal flow velocity $U_x$). It takes $\approx 0.5$~mm for $N_2$ to decrease from $10\,$\% (chamber 1) to strictly $0$ (dopant confinement). 

Dopant confinement is thus ensured for equal pressures or a slight positive gradient $\Delta{} p$. Setting $\Delta {p} = 0$~mbar provides a clear separation of gases in both chambers, with original mix and pure $He$ remaining respectively in chamber 1 and 2, while positive gradients induce $He$ leaks into chamber 1. The shortest $c_{N_2}$ transition from $10\,$\% to $0$ occurs for equal pressures. Working with $10\,$\% dopant is a dimensioning case and results are valid for lower concentrations.

We observe in simulations that increasing the working pressure from $10\,$mbar to $120\,$mbar makes the tuning of the transition position more sensitive to the pressure difference as the central aperture conductance depends on the sum of the pressure in the two chambers. The transition length remains stable. 

Simulations have also been performed for a larger aperture up to $0.95\,$mm. As conductance is higher when increasing the central aperture diameter, it shows a higher sensitivity with $\Delta p$. This sensitivity is confirmed by the experimental results in Section~\ref{sec:experimental}. 




\section{\label{sec:test}Target test bench} 

\subsection{\label{ssec:mechanical}Experimental setup}

\begin{figure}[htbp]
        \centering
        \includegraphics[width=0.5\textwidth]{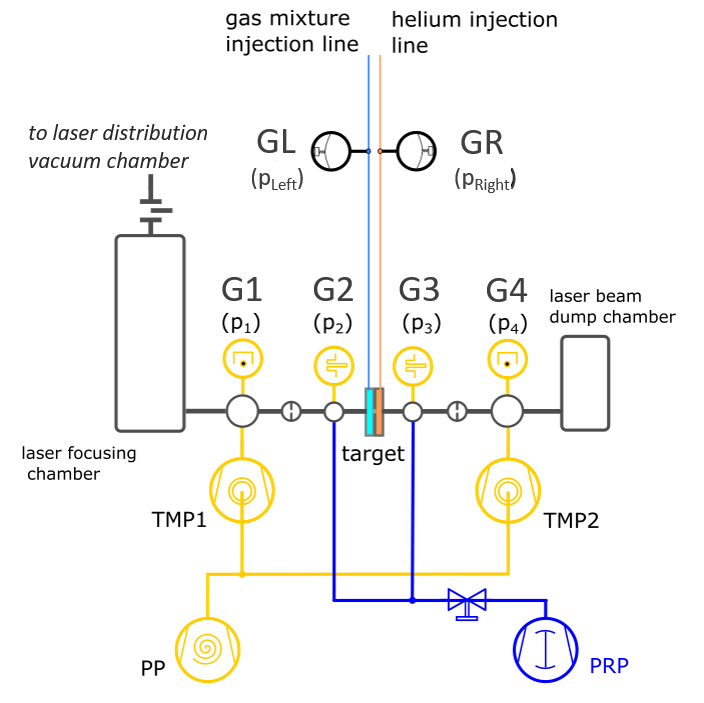}
        \caption{Schematic diagram of the vacuum setup used for cell characterisation. Pressures are monitored using gauges G1, G2, G3 and G4 respectively measuring pressures $p_1$, $p_2$, $p_3$ and $p_4$. Vacuum is ensured  by secondary turbomolecular pumps (TMP1,TMP2) with forevacuum primary pump (PP). The target gas flow is directly pumped upward and downward the target with a roots pump (PRP). Gas injection pressures are measured with ceramic piezo type gauges GL and GR respectively measuring $p_{Left}$ and $p_{Right}$.}
        \label{fig:plasmaCell_mechanical_V2}
    \end{figure}

The vacuum and mechanical setup used at IJCLab target test facility is presented in Fig.~\ref{fig:plasmaCell_mechanical_V2} with its characteristics summed-up in Tab.~\ref{tab:vacsystem}. 
    \begin{table}[htbp]
        \centering
        \begin{tabular}{|l|c|c|c|}
            \hline
             element &  parameter & value  & unit \\
             \hline
             TMP1,2 & pumping speed & $300$ & l/s \\
             TMP1,2 & gas through output & $160$ & mbar.l/s \\ 
             PP & pumping speed & $40$   & l/s \\
             PRP & pumping speed & $>400$ & l/s \\
             G1,G4 & gauge type  & cold cathode & - \\
             G2,G3 & gauge type & pirani & - \\ 
             GL,GR & gauge type & capacitance & - \\
             \hline
        \end{tabular}
        \caption{Characteristics of vacuum system elements given for helium at the working pressure}
        \label{tab:vacsystem}
    \end{table}    

\begin{figure}[htbp]
        \centering
        \includegraphics[width=0.5\textwidth]{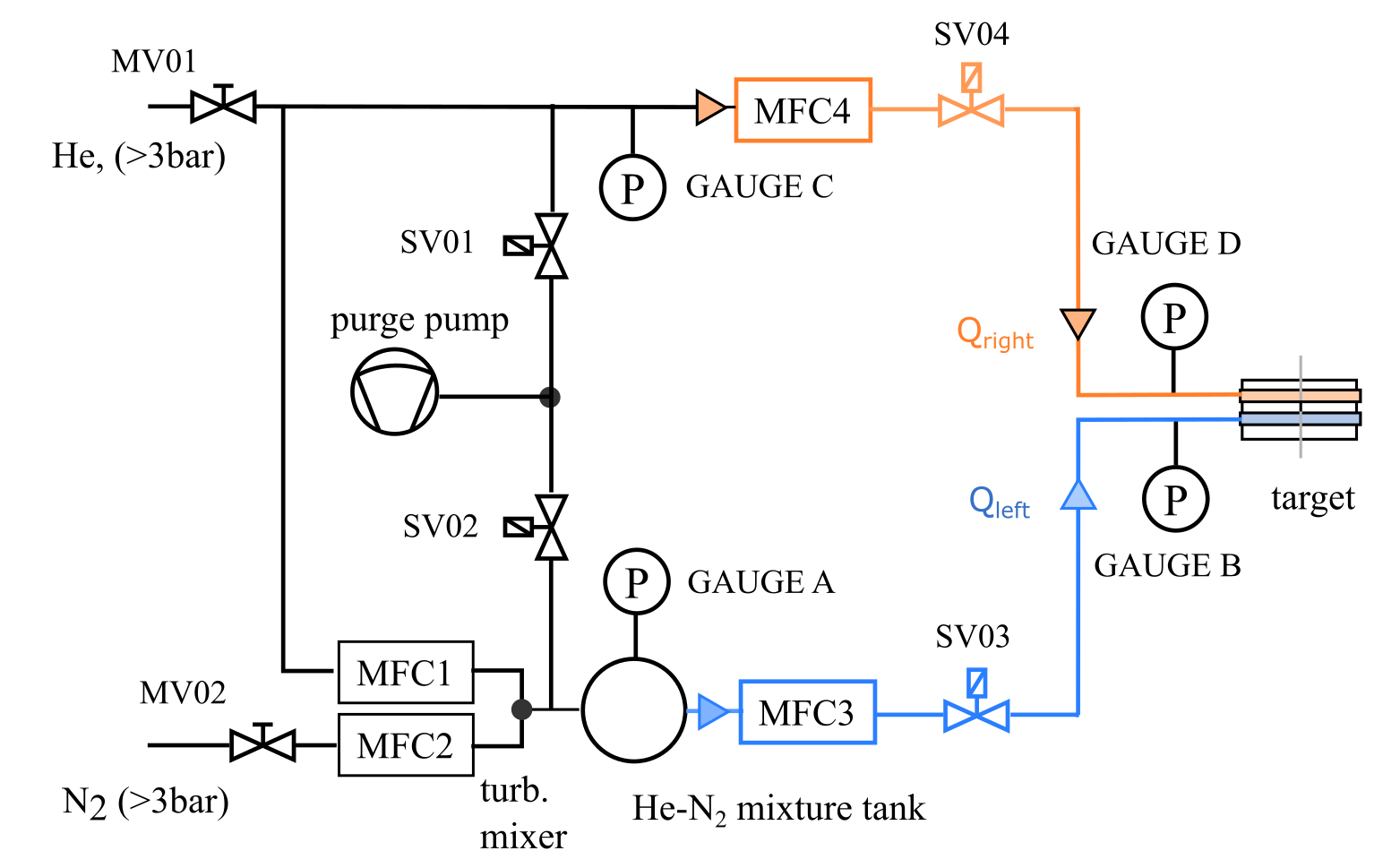}
        \caption{Schematic diagram of the gas injection system. The mass flow controllers MFC1 and MFC2 set the concentration of $N_2$ in the gas mixture. The MFC3 and MFC4 set the mass flow injected on the chamber 1 and chamber 2 respectively. The GAUGE-$A-D$ are the gauges for injection automation. SV0X are solenoid valves and MVOX manual valves.}
        \label{fig:gasInjectionSystem}
    \end{figure}

Gases are injected using a specific gas injection system (Fig.~\ref{fig:gasInjectionSystem}), with one injection line for each chamber. For chamber 1, the user specifies the $He/N_2$ mixture injection mass flow and dopant concentration in $\%$ (partial pressure ratio) within $[0;100]\pm0.2$. For chamber 2, a pure $He$ injection mass flow is set.

Several gauges monitor the pressure. Their names follow the laser propagation direction: G1, G2, G3, G4 respectively measuring $p_1$ (secondary vacuum), $p_2$ (primary vacuum), $p_3$ (primary vacuum) and $p_4$ (secondary vacuum). Pressure at the injection is monitored with capacitance gauges GL and GR, whose measurements are independent of gas type. 

Monitoring the pressure is of particular importance to constantly check the state of the cell and alert on associated pollution propagating upstream the laser line. Additionally, gauges can serve as verification tool for fluid simulations. They allow to cross-check aperture-induced pressure drop far from axis (static pressure difference between chamber 2 and downstream pumping tee for instance). This serves as flow modeling validation regarding gas thermophysical properties and flow regime (laminar VS turbulent). 

\subsection{\label{ssec:main}Laser line}

\begin{figure}[htbp]
        \centering
        \includegraphics[width=0.5\textwidth]{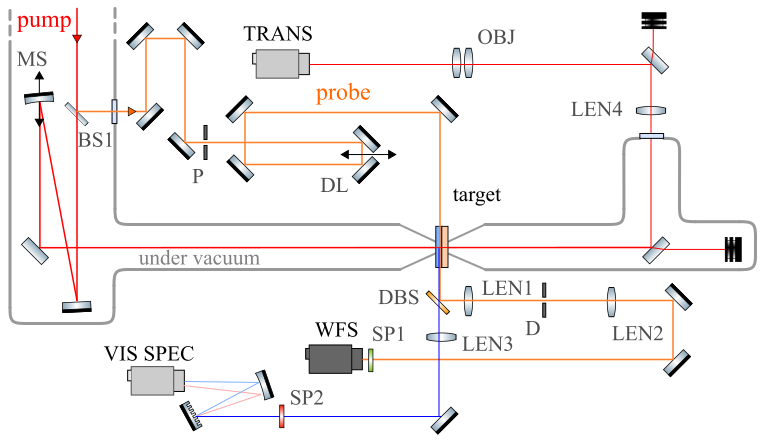}
        \caption{
        Optical scheme used for target characterisation. The optical paths for the pump and probe beams respectively are in red and orange. The pump beam is focused into the target by MS (spherical mirror), the probe beam is extracted by BS1 (5/95 beamsplitter). Other optical components are: P (pinhole), DL (motorised delay line), DBS (dichroic beamspliter), LEN1-4 (lenses), D (adjustable diaphragm), OBJ (microscope objective). Optical diagnostics are: WFS (wavefront sensor), VIS SPEC (visible spectrometer), TRANS (camera).
        }
        \label{fig:plasmaCell_optical}
    \end{figure}

The pump/probe optical setup is described in Fig.~\ref{fig:plasmaCell_optical}, where the laser beam comes from LaseriX platform \cite{laserix}. 
The pump beam (characteristics in Tab.~\ref{tab:laser_parameters}) is focused in the target and serves for ionisation. $10\,$\% of the total energy is dedicated to the probe beam.

\begin{table*}[]
    \centering
    \begin{tabular}{|l|c|c|c|}
         \hline
         Parameters & value & typical errors & unit \\
         \hline
         central wavelength; $\lambda_{0}$ & $810$ &  $\pm 1$ & nm \\ 
         minimum pulse duration (FWHM); $\tau$ &  50 & $\pm 5$ & fs \\
         repetition rate & $10$ & - & Hz \\
         Flattened Gaussian beam order; $N$ & $5$ & - & - \\
         energy on target; $E_0$ & $1 \rightarrow 60$ & $\pm 5 $ & mJ \\ 
         focal length; $f$ & $1100$ & - & mm \\
         waist in the focal plane; $w_0$ & $55$ & $\pm 5$ & $\mu$m \\
         Strehl ratio &  $0.55$ & $\pm 0.05$ &  - \\
         focal spot longitudinal position range; $\Delta x_{foc}$ & $30$ & - &  mm \\ 
         probe delay; $\Delta t$ & $180$ & $\pm 0.03$ & ps \\     
         
        \hline
    \end{tabular}
    \caption{Parameters of the laser pulse from LaseriX platform \cite{laserix} used for target gas ionisation (pump beam) and transverse optical diagnostics (probe beam) on IJCLab target characterisation test bench.}
    \label{tab:laser_parameters}
\end{table*}

\noindent The optical diagnostics used on the test bench for the target characterisation are: a wavefront sensor \cite{phasics} for plasma channel density measurement \cite{plateau2010wavefront, brandi2019optical} and a visible spectrometer \cite{femtoeasy} (or a camera \cite{basler}) for ion species measurement.



\section{\label{sec:experimental}Experimental qualification} 

\subsection{\label{ssec:density} Neutral gas pressure measurement}

Simulation are cross-calibrated with experimental results: experimental pressure measurements validate the flow hypothesis (laminar versus turbulent) for the solver and its  ability to reproduce pressure drops induced by the apertures. Simulations were performed with \textit{rhoPimpleFoam} in laminar mode.
Whether for $He$ or $N_2$, the pressure in chamber 1 is predicted with an error below $10\,$\%, which diminishes for higher injection pressures. The primary vacuum pressure $p_{3}$ prediction slightly deviates from the experiment, with a maximum error corresponding to $0.1$~mbar. Sources of deviations are: a central aperture shape not perfectly modeled, a pumping system overestimated in simulation at low pressures and the inability of the solver to model quasi-discontinuous flows for very low pressures. Simulations manage to correctly reproduce the flow down to $0.1$~mbar. The relevance of gas property choice, such as viscosity is confirmed, together with the use of laminar mode: turbulence does not have to be activated, which greatly reduces the simulation time.

\subsection{\label{ssec:edensity} Electron plasma density profiles}

\begin{figure}[htbp]
        \centering
        \includegraphics[width=0.5\textwidth]{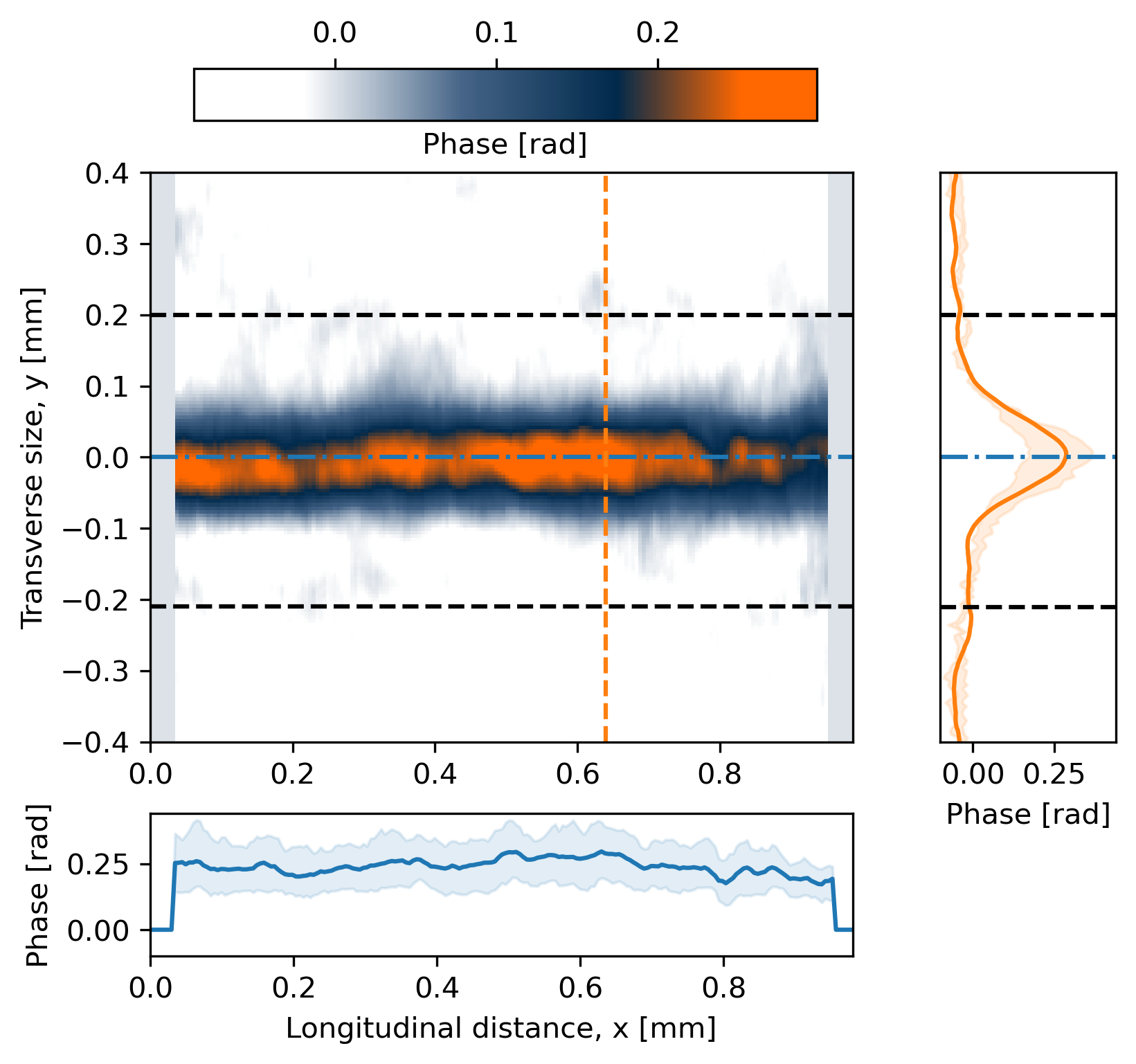}
        \caption{Averaged phase difference measured with the wavefront sensor (on $20$ consecutive shots) in chamber 1 for $He$ at $30$~mbar (pressure gauge measurement) with two additional slices: longitudinal plot extracted at $y=0$ (bottom graph), transverse plot at $x=0.64$~mm (right graph). The 'std' for each 1D plot is added in the cloud around the mean curve (computed using the shot-to-shot variation). Laser goes from left to right.
        }
        \label{fig:phasics_phase_He_30mbar}
    \end{figure}

Additionally to gauge pressure measurements far from axis, the density profile on axis is assessed using a wavefront sensor. The latter is used to record the phase difference introduced by the plasma channel in chamber 1, close to the axis. A typical phase map is presented in Fig.~\ref{fig:phasics_phase_He_30mbar}.

The plasma channel has a constant diameter of $100$~$\mu$m compatible with the laser width $2\times w_{0} = 110$~$\mu$m. 
1D plots envelopes are standard deviations computed from 20 shot series taken at $1\,$Hz \footnote{Although the pump and probe lasers are at $10\,$Hz, acquisitions are performed at $1$~Hz (software constraints).}. 
\noindent The important noise level can be explained by: ambient air density variations integrated over the whole probe beam path (a few meters), test bench vibrations or laser ablated particles projecting impurities in the chambers. Since the phase remains quite stable above $0.2$~mm from axis (no plasma), cropping is performed on each phase map, to increase the signal-to-noise ratio. In the worst case (low pressure), the signal-to-noise ratio was always $>4$. 

\begin{figure}[htbp]
        \centering
        \includegraphics[width=0.5\textwidth]{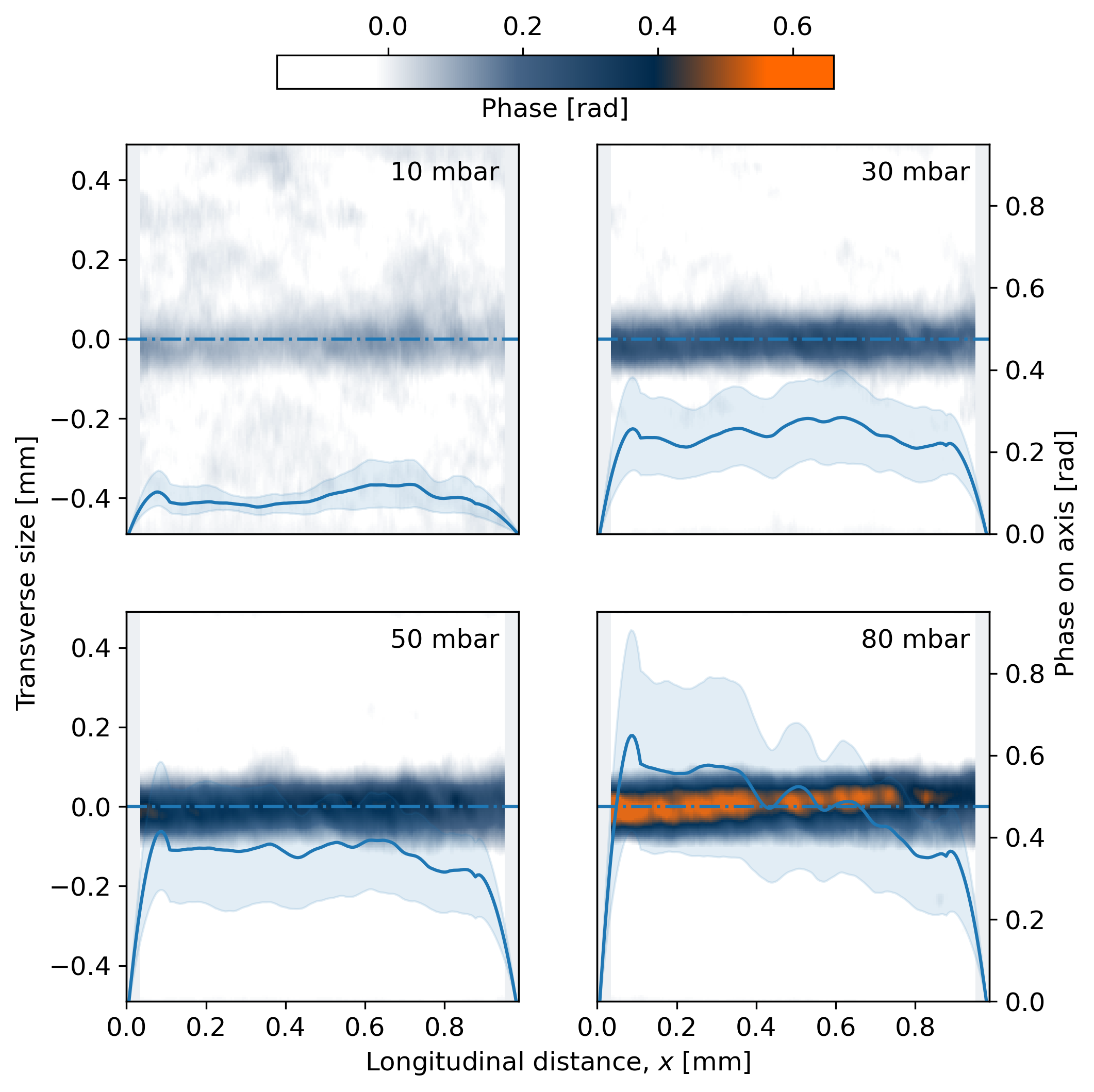}
        \caption{Averaged phases (on $20$ consecutive shots) acquired with the wavefront sensor for $He$ in chamber 1 at $10$, $30$, $50$ and $80$~mbar. Longitudinal plots (blue line)
        extracted at $y=0$ are added on each image, with corresponding 'std' (blue filled) based on their shot-to-shot variation. Laser goes from left to right.
        }
        \label{fig:phasics_phases_He}
    \end{figure}

Typical phase maps acquired for $He$ at $10$, $30$, $50$ and $80$~mbar in chamber 1 are presented in Fig.~\ref{fig:phasics_phases_He}.
They display similar features than for Fig.~\ref{fig:phasics_phase_He_30mbar}. However, for a high pressure ($80$~mbar) a longitudinal gradient appears, probably due to ionisation defocusing of the laser. 

\begin{figure}[htbp]
        \centering
        \includegraphics[width=0.5\textwidth]{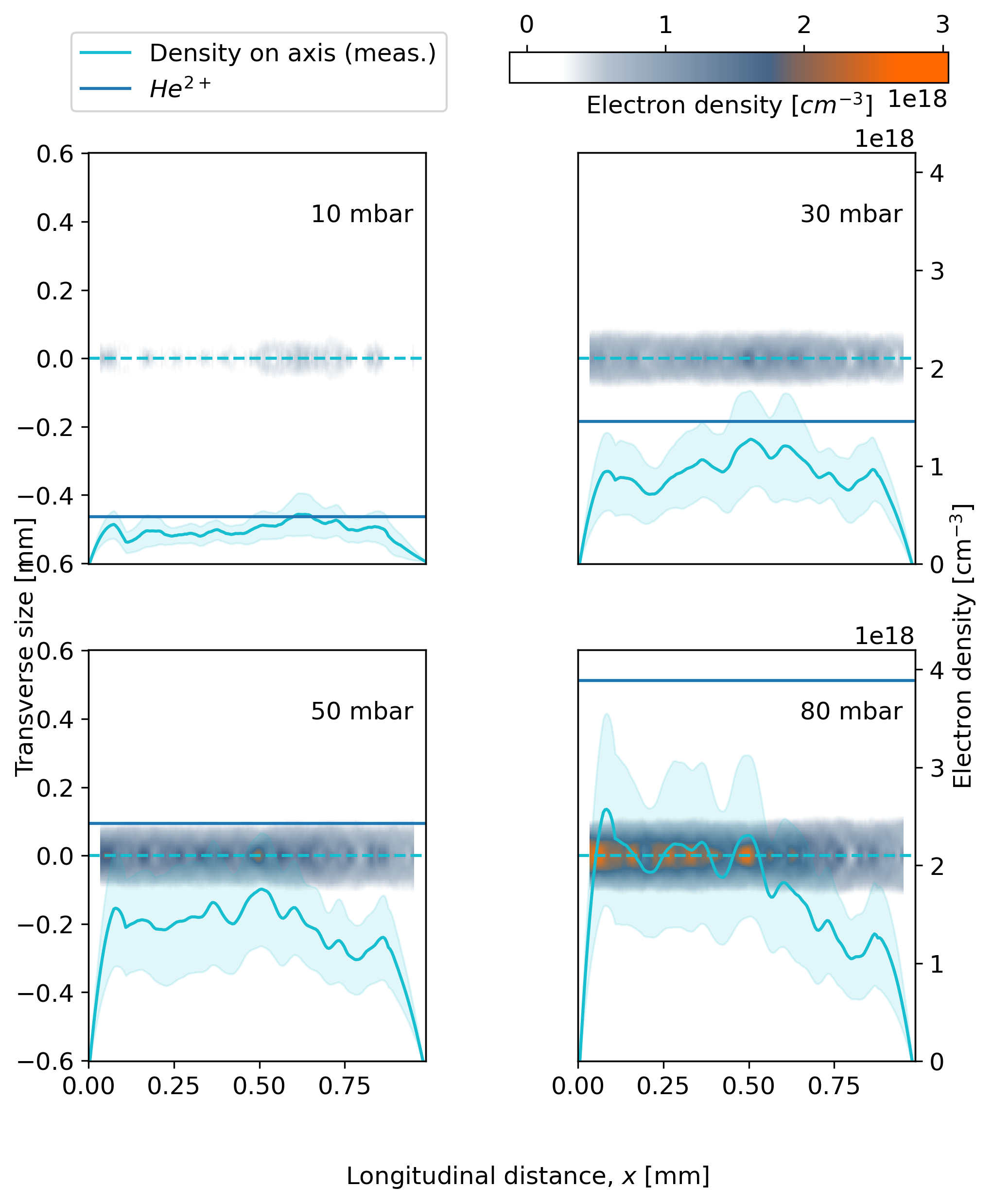}
        \caption{Electron densities obtained with Abel inversion 
        on Fig.~\ref{fig:phasics_phases_He} phases maps. Additional symmetrisation is applied. A longitudinal plot (light blue line) at $y=0$ is added.
        The straight horizontal line (dark blue) represents the density corresponding to fully ionised $He$. Laser goes from left to right.
        }
        \label{fig:phasics_densities_He}
    \end{figure}

Abel inversion is used on phase maps to retrieve the corresponding electron plasma density distribution.
The resulting density maps are presented in Fig.~\ref{fig:phasics_densities_He} were inversion has been performed on phase maps from Fig.~\ref{fig:phasics_phases_He}. 

Theoretical maximum densities expected for fully ionised $He$ at $10$, $30$, $50$ and $80$~mbar are respectively $4.86 \times 10^{17}$, $1.46 \times 10^{18}$, $2.43 \times 10^{18}$ and $3.89 \times 10^{18}$~cm${^{-3}}$. They are indicated with horizontal lines on Fig.\ref{fig:phasics_densities_He}. For low pressure, electron density on axis is quite constant, slightly below the theoretical value with some peaks along propagation. 
For $50$ and $80~$~mbar, ionisation likely remains at the $He^{+}$ level, with a progressive drop for $80$~mbar above $0.55$~mm. This confirms that at high pressure, the pump beam is not intense enough to ionise the two levels of $He$, due to stronger ionisation defocusing.


We conclude that theoretical measured pressures match simulations, with a longitudinal density having a constant plateau-like shape in the first chamber. A similar behaviour is expected for chamber 2, since its geometry is quite similar to chamber 1. 

\subsection{\label{ssec:dopant_exp}Dopant longitudinal profile}


The prototype ability to confine the dopant is experimentally assessed with an imaging spectrometer. It relies on excited species emission, with a
resolution of $1\,$nm over the considered wavelength range ($400-600\,$nm). 

\begin{figure}[htbp]
        \centering
        \includegraphics[width=0.5\textwidth]{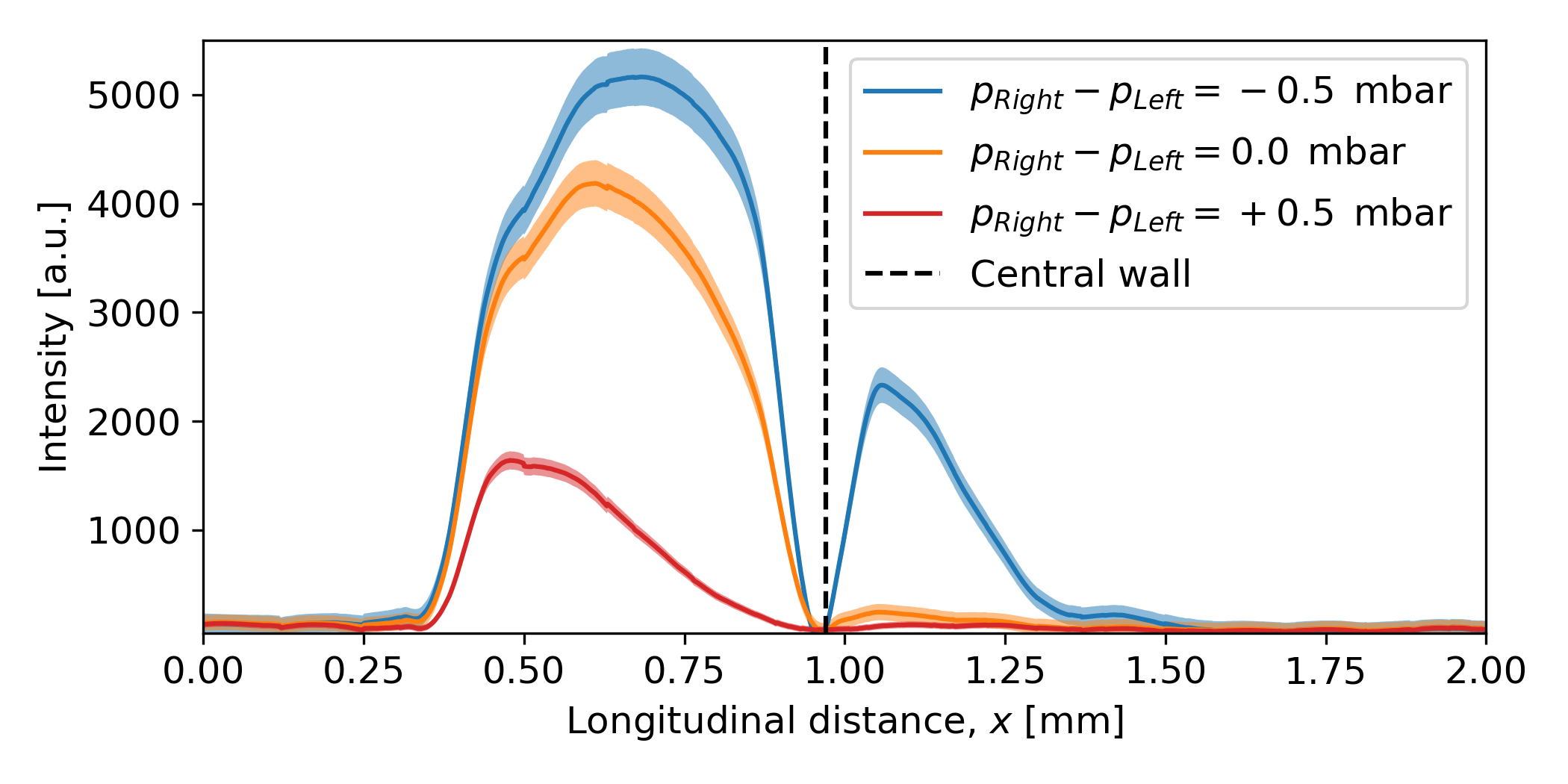}
        \caption{Dopant localisation using spectrometer measurements for different gradient values $\Delta p = p_{Right} - p_{Left}$ between chamber 1 and chamber 2, with an average plateau-pressure on axis of $30$~mbar. Pure $N_2$ and pure $He$ respectively are injected in chamber 1 and 2.
        Geometry is $(1.5,1.15,0.25,0.95,1.5,0.5,0.95,0.5)$. The central aperture (central wall) position is added (black dashed line). Laser goes from left to right.
        }
        \label{fig:dopant_spectro}
    \end{figure}

To get a good resolution on the diffusion of $N_2$ along the two chambers, pure $N_2$ is injected in chamber 1, while pure $He$ is injected in chamber 2. This case is not representative of a typical working point for the target, but will provide conservative information on dopant confinement. The spectrometer gives the possibility to select various emission lines to track simultaneously the corresponding species. Experimental results for different pressure gradients $\Delta p = p_{Right} - p_{Left}$ are presented in Fig.~\ref{fig:dopant_spectro}, with the largest achievable central aperture diameter ($0.95$~mm).


For $p_{Left} = p_{Right}$, the dopant is confined in chamber 1 (orange curve in Fig.~\ref{fig:dopant_spectro}). For a slight positive gradient, the dopant is strongly pushed to the left (red curve in Fig.~\ref{fig:dopant_spectro}), reducing the size of the injection zone (truncated ionisation injection). On the contrary, even for a rather small negative $\Delta p$, the dopant leaks into chamber 2 (blue curve in Fig.~\ref{fig:dopant_spectro}). Nevertheless, a clear stable confinement of the gas mixture is demonstrated in agreement with the simulations.

\subsection{\label{ssec:lifetime}Target lifetime}


Previous characterisation and simulations of course remain valid as long as the target retains its geometry under high intensity laser irradiation.

The target main body is composed of aluminium, while nozzles are either in aluminium or in ceramics (MACOR). The most critical part of the design is the inlet nozzle. As shown in numerical simulations and experimental measurements the gas mixture confinement can be obtained with a central wall aperture diameter up to $\approx 1\,$mm.

\begin{figure}[htbp]
        \centering
        \includegraphics[width=0.5\textwidth]{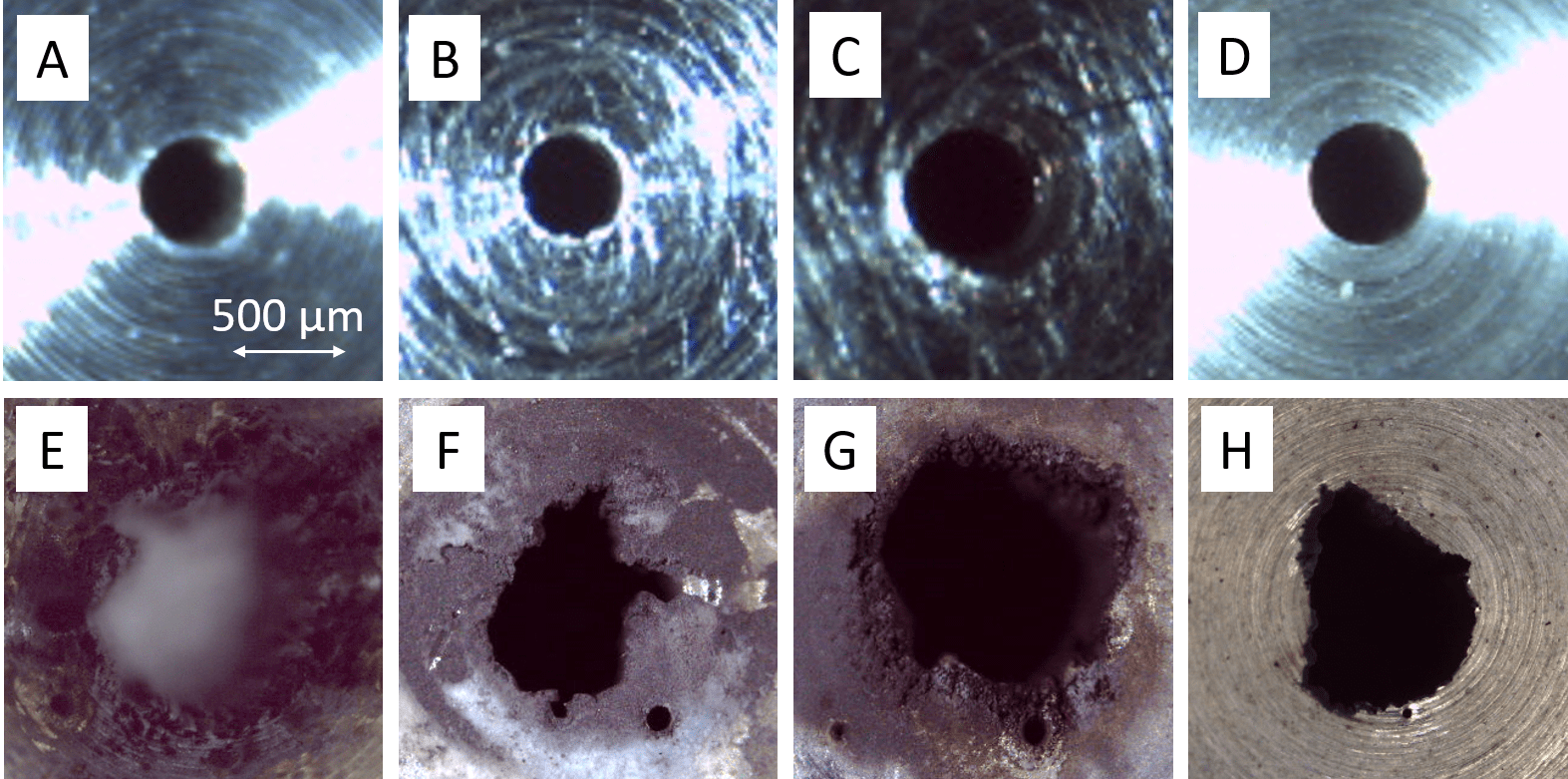}
        \caption{Aluminium nozzle evolution before (top row) and after (bottom row) $300\,000$~shots at $60\,$mJ. Images are: inlet nozzle concave face (A,E), inlet nozzle convex face (B,F), outlet nozzle convex face (C,G), outlet nozzle concave face (D,H). Initial nozzles dimensions are $D_1 = 520\pm10$~$\mu$m (inlet nozzle, average of 'A' and 'B') and $D_5 = 600\pm10$~$\mu$m (outlet nozzle, average of 'C' and 'D'). Damaged nozzles dimensions are $D_1 = 910\pm10$~$\mu$m (inlet nozzle, average of 'E' and 'F') and $D_5 = 990\pm10$~$\mu$m (outlet nozzle, average of 'G' and 'H'), with damaged apertures approximated as circles. Nozzle aperture lengths are $L_{1}=1$~mm (A,B,E,F) and $L_{5}=3$~mm (C,D,G,H).}
        \label{fig:nozzles_B2_before_after}
    \end{figure}

\noindent Typical aperture dimensions and shape variations before and after $300\,000$~shots at $60$~mJ for aluminium nozzles are presented in Fig.~\ref{fig:nozzles_B2_before_after}.
Our experimental observation is that even at $60$~mJ (well below the $1$~J required for laser-plasma acceleration), the aluminium nozzles are strongly damaged.

\begin{figure}[htbp]
        \centering
        \includegraphics[width=0.5\textwidth]{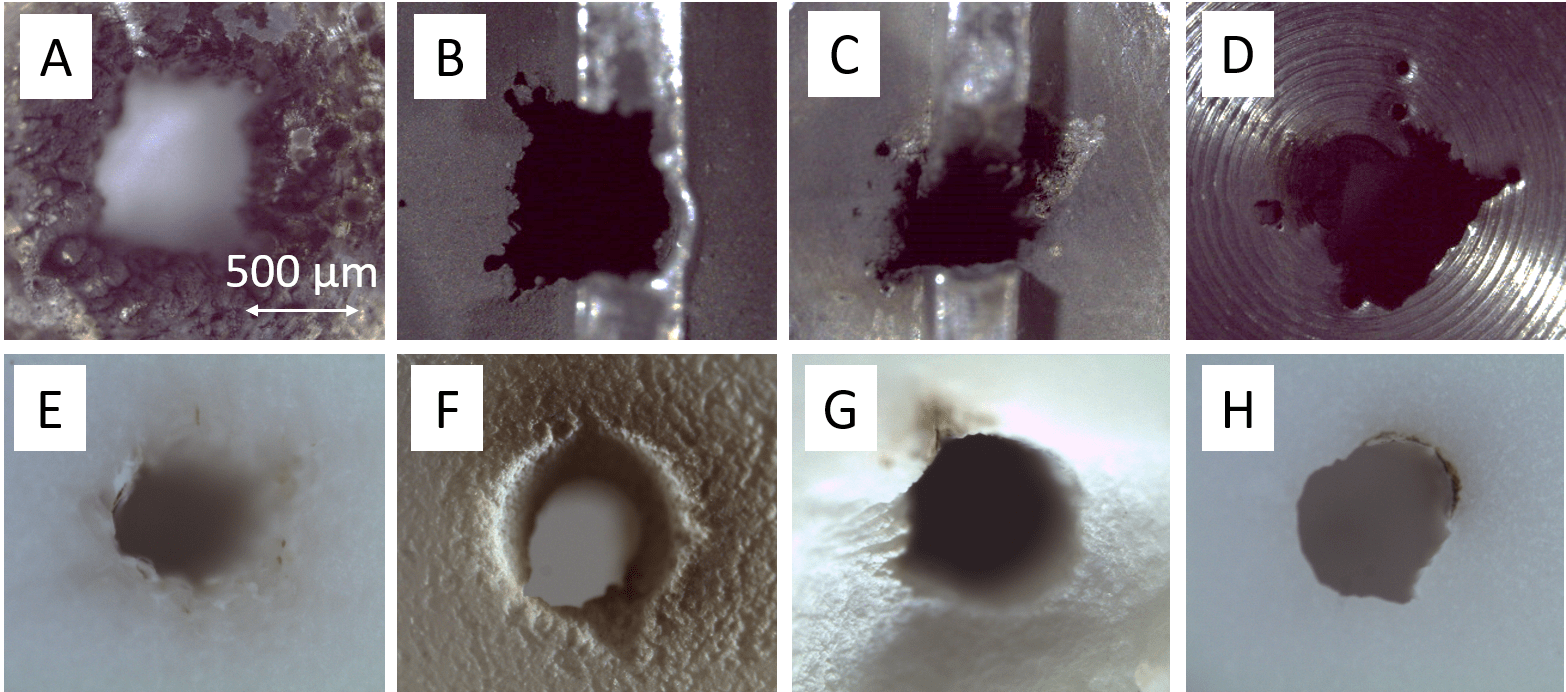}
        \caption{Comparison between aluminium (top row) and MACOR nozzles (bottom row) after approximately $300\,000$ shots at $60\,$mJ. Images are: inlet nozzle concave face (A,E), inlet nozzle convex face (B,F), outlet nozzle convex face (C,G), outlet nozzle concave face (D,H). Initial diameters were $D_{1}=D_{5}=500\pm10$~$\mu$m for all four nozzles. Damaged apertures are approximated as circles. Damaged aluminium nozzles dimensions are $D_1 = 830\pm10$~$\mu$m (inlet nozzle, average of 'A' and 'B') and $D_5 = 740\pm10$~$\mu$m (outlet nozzle, average of 'C' and 'D'). Damaged MACOR nozzles dimensions are $D_1 = 710\pm10$~$\mu$m (inlet nozzle, average of 'E' and 'F') and $D_5 = 740\pm10$~$\mu$m (outlet nozzle, average of 'G' and 'G'). Nozzle aperture lengths are $L_{1}=1.5$~mm (A,B,E,F) and $L_{5}=1.5$~mm (C,D,G,H).}
        \label{fig:nozzles_comparison_alu_macor}
    \end{figure}

A solution is to use MACOR nozzles as shown in Fig.~\ref{fig:nozzles_comparison_alu_macor}.
Qualitatively, higher MACOR resistance is visible on the post-mortem pictures.

An online estimation of the nozzle state can be done through pressure control. We experimentally observed an inlet pumping tee pressure rise up to the mbar range for aluminium nozzles after $\sim 30$~min of operation, while remaining in a $10^{-1}\,$mbar range for MACOR nozzles. Ceramics greatly improve the cell lifetime. This conclusion from the characterisation test bench has to be confirmed on real scale laser plasma experiments.



In the case of acceptable nozzle deterioration, we are able to take into account the evolution of nozzle apertures with time, with continuous adjustment of the gas injection flows to maintain a constant pressure in the chambers. In the optimisation of electron beam parameters, the laser focusing position may also be tuned during operation to counterbalance the elongation of the in-ramp length. After a few thousand shots, a saturation of the ablation is observed, leading to more stability.

Regarding optical diagnostic, the design is quite robust and for more than $10^6$~shots at $60$~mJ, no optical window had to be replaced or even cleaned, allowing continuous cell characterisation and monitoring through transverse diagnostics. This is favoured by the distance between plasma and window of roughly $30$~mm, preventing direct deposition of ablated material. 

\section{\label{sec:discussion}Discussion and conclusion}

The density profile of a two-chamber gas cell prototype for ionisation injection has been assessed using the open-source fluid simulation library OpenFOAM. It has been cross-checked with experimental results comming from the diagnostics installed on the LaseriX test bench. Simulation results are open to the scientific community. 

Our multi-cell target design offers density distribution control and precise dopant confinement, which have been experimentally demonstrated with online diagnostics that also allow to monitor the target state evolution during the experiment.

For emittance conservation issues, the output nozzle can be shaped to optimise passive plasma lensing with an adapted density out-ramp. 
The target integration in the beamline also offers a compact laser-plasma injector design. This allows for a compact beam transport line for further injection into a second accelerating stage. For example, the first magnet for PALLAS project can theoretically be put as close as $\approx 15\,$cm from the source.


The results of this target design characterisation have been used as input for Particle-in-Cell simulations, with the aim to find optimal working points for electron injection. Four parameters have been varied: chamber pressure $p_{Left}$ (with $p_{Left} = p_{Right}$), dopant concentration $c_{N_2}$, laser energy $E_0$ and laser focal position $x_{foc}$ \cite{drobniak2023}. The numerical results show that electron beams with a charge over $30$~pC, energy ranging between $150-250$~MeV, energy spread below $5\,\%$ and transverse normalised emittance below $2\,\mu{}$m can be obtained.



\appendix

\section{Appendixes}

\bibliography{biblio.bib}

\end{document}